\documentclass[letterpaper,journal]{IEEEtran}

\usepackage{amsmath,amsfonts}
\usepackage{algorithm}
\usepackage{algpseudocode}
\usepackage[caption=false,font=normalsize,labelfont=rm,textfont=rm]{subfig}
\usepackage{textcomp}
\usepackage{stfloats}
\usepackage{url}
\usepackage{verbatim}
\usepackage{graphicx}
\usepackage{array}
\usepackage{booktabs}
\usepackage{cite}
\usepackage{xcolor}
\usepackage{mathtools}
\usepackage{bm}
\usepackage{amsthm}
\usepackage{float}
\usepackage[export]{adjustbox}
\usepackage[
    colorlinks=true,
    linkcolor=blue,
    citecolor=blue,
    urlcolor=blue
]{hyperref}

\newcommand{\R}{\mathbb{R}}                     

\newcommand{\tb}{\mathrm{bat}}

\newcommand{\ts}{\mathrm{sc}}

\usepackage{xcolor}

\newcommand{\tg}{\mathrm{g}}

\begin{document}

\title{Grid-Mode-Aware Model Predictive Control of Hybrid Energy Storage Systems for AI Data Center Power Smoothing}

\author{Xin Chen,~\IEEEmembership{Member,~IEEE}

\thanks{X. Chen is with the Department of Electrical and Computer Engineering, Texas A\&M University, USA. Email: xin\_chen@tamu.edu.} 
	
\thanks{ 
The work was supported in part by NSF CAREER 2541998, in part by the Power Systems Engineering Research Center (PSERC), and in part by the Consortium on AI and Large Flexible Load (CALL) at Texas A\&M University. (\emph{Corresponding author: Xin Chen}). 
} 

}





\maketitle

\begin{abstract}

To facilitate the grid-friendly integration of highly variable AI data center loads, this paper proposes a grid-mode-aware model predictive control (G-MPC) framework for managing a hybrid energy storage system (HESS) to smooth grid-side power demand. The framework optimally coordinates a battery energy storage system (BESS) and a supercapacitor (SC) by solving a multi-step optimization problem in a receding-horizon manner. In particular, band-pass filter dynamics are directly embedded in the G-MPC formulation to extract and suppress grid-side power components associated with vulnerable grid oscillatory modes, thus mitigating load-induced grid oscillations. The resulting G-MPC optimization jointly minimizes violations of grid-side power-envelope, ramp-rate, and modal-power requirements and the degradation and power-ramping costs of the BESS and SC, while satisfying power limits, state-of-charge limits, and other operational constraints. To enable real-time implementation, a fix-and-re-optimize algorithm is developed to solve each G-MPC problem efficiently while preventing simultaneous charging and discharging. Extensive simulations demonstrate the effectiveness, flexibility, and computational efficiency of the proposed framework. The results also highlight the importance of explicitly suppressing power components associated with vulnerable grid modes, rather than merely reducing overall load variations, to effectively mitigate grid oscillations.

\end{abstract}

\begin{IEEEkeywords}
AI data center, power smoothing, hybrid energy storage, grid mode awareness, model predictive control.
\end{IEEEkeywords}

\section{Introduction}
\label{sec:introduction}


\IEEEPARstart{I}{n} recent years, the rapid advances in artificial intelligence (AI) are driving an unprecedented expansion of data center infrastructure \cite{iea2024energyai,chen2025electricity}. 
In addition to their substantial electricity demand, AI data centers exhibit fast and cyclic power variations that differ fundamentally from those of conventional loads. Specifically,
during the training of large AI models, hundreds of thousands of GPUs operate synchronously and transition repeatedly among computation-intensive, communication, and checkpointing phases, resulting in facility-level power swings ranging from tens to hundreds of megawatts \cite{choukse2025power}.
As a consequence, these large and rapid power fluctuations pose significant challenges to real-time power balancing and grid frequency stability. More importantly, periodic or quasi-periodic load variations can act as sustained forcing sources that excite and amplify poorly damped local or inter-area oscillatory modes \cite{ko2026widearea,valverde2025forced}. 
As reported by the North American Electric Reliability Corporation (NERC) \cite{nerc2025largeloads}, large data center loads may destabilize existing power system oscillatory modes and 
induce forced subsynchronous oscillations that can cause torsional damage to generators and threaten grid stability.
Therefore, grid operators and utilities have begun developing grid codes or interconnection requirements to regulate the load behavior of large AI data centers. For instance, recent NERC guidance \cite{nerc2026risk} recommends interconnection requirements such as load-variation amplitude thresholds, operational ramp-rate limits, oscillation attenuation metrics, limits on the amplitude and frequency of oscillatory demand, etc.

To comply with these requirements, behind-the-meter energy storage offers AI data centers an effective way to reshape highly variable  demand into a more grid-friendly power profile \cite{rahman2026energy,mohammadi2026grid}. However, a single energy storage technology generally cannot simultaneously satisfy the wide range of power, energy, response-speed, lifetime, and cost requirements to cope with AI workload variations \cite{chen2025electricity,mohammadi2026grid}. 
A battery energy storage system (BESS) provides high energy density for compensating sustained power variations, but frequent high-power cycling can accelerate battery degradation and require costly oversizing. In contrast, a supercapacitor (SC) offers high power density, fast response, and an extremely long cycle life on the order of millions of cycles,  
while its low energy density and high capital cost limit its ability to support sustained power deviations \cite{durvasulu2023technology}.
Hence, a hybrid energy storage system (HESS) combining a BESS and an SC can leverage their complementary strengths: the BESS compensates for energy-dominant, relatively slow variations, whereas the SC absorbs power-dominant, fast, and frequent variations. 
In this way, the HESS can reduce battery cycling stress while providing the power and energy capabilities required for AI data center power smoothing \cite{liu2015heb,zheng2017hybrid}.

Several studies have investigated the coordinated use of BESS and SC for data center load management. In \cite{liu2015heb}, a hybrid energy-buffering architecture and a real system prototype are developed to mitigate power mismatches in data centers with improved energy efficiency and economy.
Reference \cite{zheng2017hybrid} quantitatively compares the long-term operational costs of three different energy
storage options: battery only, SC only, and a hybrid battery-SC solution for data center power shaving and capping. In \cite{abera2026battery}, a battery control framework is proposed for regulating AI data center power ramp rates, with closed-form control laws derived by leveraging fast online change-point detection and the underlying optimal control problem structure. 
Recent work \cite{ko2025mitigation} proposes 
a high-pass filter scheme that separates data center demand into slow- and fast-varying components, which are then allocated to the BESS and SC, respectively, through coordinated multi-timescale PID control. In \cite{you2026source}, 
a similar fixed frequency-decomposition rule is used to allocate data center load variations between the BESS and SC, while an offline-trained differentiable predictive control policy provides finite-horizon corrections with a one-step constraint safeguard.

Nevertheless, existing HESS control methods retain several key limitations. First, 
frequency-decomposition-based schemes simply rely on a predetermined cutoff frequency to allocate power variations between the BESS and SC, without explicitly accounting for HESS operating costs or physical constraints. Second, in practice, the objective is not to completely eliminate all data center power fluctuations, but to comply with grid-side standards and performance requirements; however, existing methods may lack the flexibility to explicitly accommodate various requirements, including power variation thresholds, ramp rate limits, oscillation attenuation criteria, etc. 
Third, and more importantly, effective data center power smoothing should account for grid oscillatory modes; because the grid impact of load variations depends not only on their magnitude and ramp rates but also on their frequency components and proximity to poorly damped modes. As a result, a smoother load profile can produce larger grid-side oscillations if it contains components near vulnerable modal frequencies, whereas larger off-modal variations may have a limited oscillatory impact. 
This \emph{grid-mode awareness} is generally not considered in existing HESS control methods, yet it is particularly important when HESS power and energy capacities are limited, as it prioritizes energy storage resources on suppressing the most critical power components. Otherwise, storage capacity may be expended on dynamically benign variations, leaving the components that strongly excite the grid insufficiently attenuated. 
Recent NERC guidance also suggests establishing amplitude and frequency limits on oscillatory demand from large data center loads and avoiding natural frequencies identified through system planning studies \cite{nerc2026risk}.

Motivated by these challenges, this paper develops a \emph{grid-mode-aware model predictive control} (G-MPC) framework for coordinating the BESS and SC in an AI data center to smooth its grid-side power demand. At each control step, the proposed G-MPC leverages the predicted data center demand to optimize the power trajectories of the BESS and SC over a multi-step prediction horizon. 
This receding-horizon structure enables the HESS to respond in advance to anticipated data center load transitions and to better manage its limited energy capacity over time \cite{schwenzer2021review}. 
The vulnerable components of the grid-side power are extracted using band-pass filter dynamics centered on the grid modes identified by the grid operator or through measurement-based modal estimation.
The filter dynamics are directly embedded into the MPC formulation, allowing these components to be explicitly constrained and penalized. 
The resulting optimization jointly minimizes violations of the grid-side power-envelope, ramp-rate, and modal-power requirements, as well as the BESS and SC degradation and power-ramping costs, while satisfying the charging and discharging power limits, state-of-charge (SOC) limits, and other operational constraints. Rather than allocating data center load variations between the BESS and SC through predefined frequency separation, the G-MPC framework dynamically optimizes power allocation based on their real-time operating states and distinct characteristics in degradation and power-ramping costs, power limits, and energy capacities. 

The key contributions of this paper are listed as follows:
\begin{itemize}
    \item [1)] We formulate the coordinated control of the BESS and SC for AI data center power smoothing as an optimal G-MPC framework that jointly minimizes violations of grid-side load-variation requirements and HESS operating costs, while satisfying the HESS operational constraints. Moreover, the proposed framework can be readily adapted to evolving interconnection requirements by modifying objective terms and constraints without redesigning the overall controller, thus providing a flexible interface for practical grid-data center coordination.
    \item [2)] We directly incorporate power-mode extraction dynamics into the G-MPC models and explicitly constrain and penalize critical load components near vulnerable grid oscillatory modes, enabling the grid-mode-aware optimal suppression of AI data center load variations to mitigate grid oscillations and enhance system stability.
    \item [3)] A computationally efficient \emph{fix-and-re-optimize} algorithm is developed for solving each G-MPC optimization problem through convex quadratic programming, 
    which ensures no simultaneous charging and discharging of the BESS and SC, while providing a computable upper bound on the optimality gap of the resulting control solutions.
\end{itemize}

 Moreover, extensive simulations demonstrate the effectiveness, flexibility, and computational efficiency of the proposed framework. In particular, the ablation studies show that explicit mode suppression is critical for mitigating grid oscillations induced by AI data center loads: a controller without targeted mode suppression may yield a smoother grid-side power profile yet produce larger grid frequency oscillations due to retaining a power component near a vulnerable grid mode.

The remainder of this paper is organized as follows. Section~\ref{sec:problem_setup} introduces the HESS system control and power mode extraction. Section~\ref{sec:mpc} develops the G-MPC model formulation and the solution algorithm. Section~\ref{sec:simulation} presents the simulation results. Conclusions are drawn in Section~\ref{sec:conclusion}.

\section{Problem Setup and Power Mode Extraction}
\label{sec:problem_setup}

This section introduces the problem setting that controls a HESS to smooth AI data center demand for satisfying grid-side requirements, and presents the method for extracting load components associated with vulnerable grid oscillatory modes.

\subsection{Hybrid Energy Storage System in AI Data Center}
\label{subsec:hess_setup}

\begin{figure}
    \centering
    \includegraphics[width=0.9\linewidth]{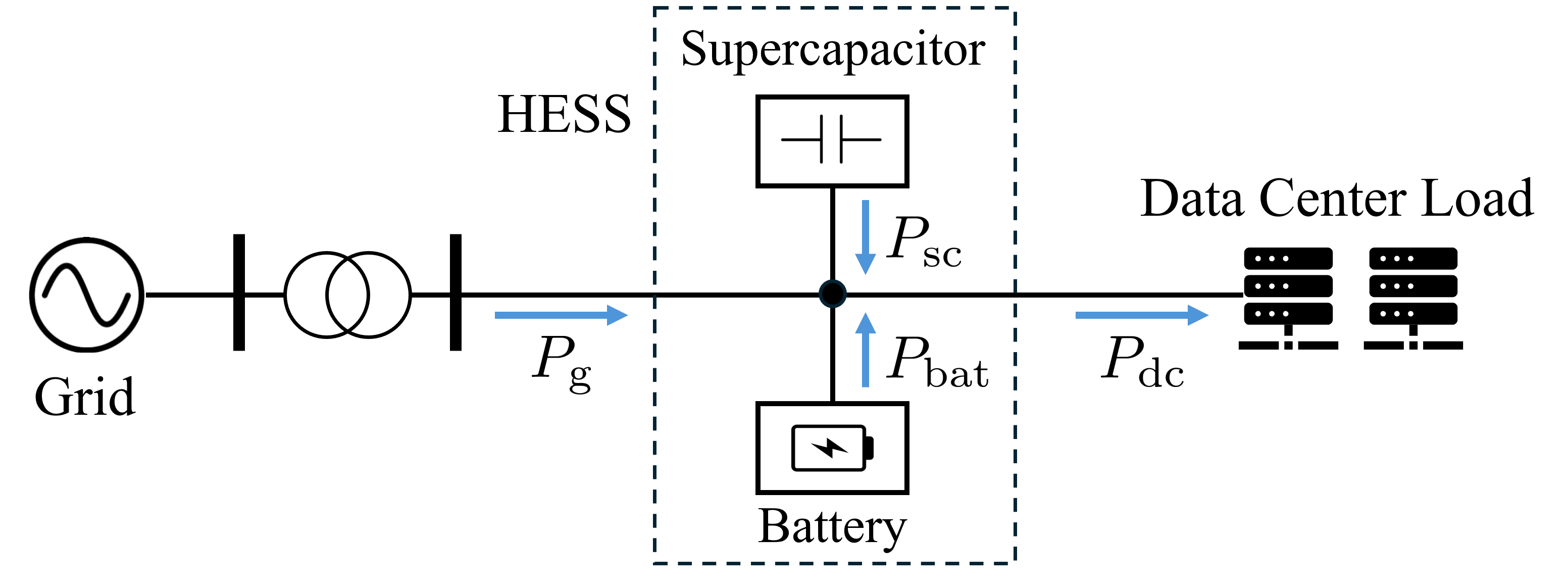}
    \caption{Structure of a grid-connected AI data center equipped with a HESS.}
    \label{fig:hess_architecture}
\end{figure}

As illustrated in Figure~\ref{fig:hess_architecture}, we
consider a grid-connected AI data center equipped with a HESS comprising a BESS and a SC. Both the BESS and SC are connected
behind the point of interconnection and are interfaced through bidirectional power converters. 
Let $P_{\mathrm{dc}}(t)$ denote the active power demand of the
AI data center at time $t$. Let $P_{\mathrm{bat}}(t)$ and $P_{\mathrm{sc}}(t)$
denote the active powers supplied by the BESS and SC, respectively;
positive (negative) power indicates discharging (charging). The active
power seen by the grid at the interconnection point is:
\begin{equation}
P_{\mathrm{g}}(t)
=
P_{\mathrm{dc}}(t)
-
P_{\mathrm{bat}}(t)
-
P_{\mathrm{sc}}(t).
\label{eq:grid_power_balance_setup}
\end{equation}
The objective of the HESS control is to reshape the highly variable AI
data center load $P_{\mathrm{dc}}(t)$ into a grid-friendly power profile $P_{\mathrm{g}}(t)$ with the desired
dynamic characteristics.


From the grid perspective, we consider three 
typical performance requirements on 
$P_{\mathrm{g}}(t)$. First, the magnitude of
the data center load should stay within an admissible power
envelope \eqref{eq:grid_power_envelope} over a specified operational period:
\begin{equation}
\underline{P}_{\mathrm{g}}
\leq
P_{\mathrm{g}}(t)
\leq
\overline{P}_{\mathrm{g}},
\label{eq:grid_power_envelope}
\end{equation}
where $\underline{P}_{\mathrm{g}}$ and $\bar{P}_{\mathrm{g}}$ are the lower and upper load limits.

Second, its upward and downward ramp rates should satisfy:
\begin{equation}
-R_{\mathrm{g}}^{\mathrm{down}}
\leq
\frac{dP_{\mathrm{g}}(t)}{dt}
\leq
R_{\mathrm{g}}^{\mathrm{up}},
\label{eq:grid_ramp_setup}
\end{equation}
where $R_{\mathrm{g}}^{\mathrm{down}}$ and $R_{\mathrm{g}}^{\mathrm{up}}$ denote the downward and upward ramping rate limits, respectively.

Third, to support grid stability, the grid-side demand $P_{\mathrm{g}}(t)$ of the AI data center should have limited 
power components at frequencies near poorly damped grid oscillatory modes. That is
because periodic or quasi-periodic power variations concentrated near a vulnerable oscillation frequency can continuously excite the corresponding mode, potentially aggravating oscillations, damaging electric
equipment, and threatening system stability. Therefore, effective power smoothing should not only limit the overall variation magnitude and ramp rate of the AI data center load, but also explicitly suppress power variations within frequency bands associated with vulnerable grid modes.
This motivates the grid-mode-aware power smoothing scheme developed in the next subsection.

\subsection{Power Mode Extraction of AI Data Center Load}
\label{subsec:grid_mode_extraction}

Let $\mathcal{M}$ denote the set of vulnerable grid oscillatory modes,
identified by the grid operator or obtained through measurement-based
modal estimation. For each mode $m\in\mathcal{M}$, let $f_m$
denote its oscillation frequency and define $\omega_m=2\pi f_m$. To quantify the oscillatory component of
the grid-side data center power $P_\tg(t)$ around  mode $m$, a second-order band-pass filter $W_m(s)$ is introduced as \eqref{eq:modal_bandpass_filter}:
\begin{align}
W_m(s) = \frac{2\zeta_{m}\omega_m s
}{s^2 +2\zeta_{m}\omega_m s+\omega_m^2},
\label{eq:modal_bandpass_filter}
\end{align}
where $\zeta_{m}>0$ is a parameter that determines the width of the frequency band of
interest, and $s$ is the Laplace variable.

The power component of the grid-side data center demand $P_\tg(t)$ associated
with mode $m$ is then extracted as:
\begin{equation}
Z_m(s)
=
W_m(s)P_\tg(s),
\label{eq:mode_filter_output_laplace}
\end{equation}
where $P_\tg(s)=\mathcal{L}\{P_\tg(t)\}$ and $\mathcal{L}\{\cdot\}$
represents the Laplace transform. 
To implement \eqref{eq:mode_filter_output_laplace} in the time domain for control design and avoid the derivative term of $P_{\mathrm{g}}(t)$, we introduce a new state $Q_m(s)=\mathcal{L}\{q_m(t)\}$ given by \eqref{eq:qm}:
\begin{align}\label{eq:qm}
    Q_m(s) \coloneqq\frac{Z_m(s)}{2\zeta_{m}\omega_m s}=\frac{1
}{s^2 +2\zeta_{m}\omega_m s+\omega_m^2}P_\tg(s).
\end{align}
Then, in the time domain, \eqref{eq:qm} leads to:
\begin{align}
&\ddot q_m
+
2\zeta_{m}\omega_m\dot q_m
+
\omega_m^2q_m
=
P_{\mathrm{g}},
\label{eq:auxiliary_mode_filter}\\
& z_m = 2\zeta_{m}\omega_m\dot q_m,
\label{eq:mode_filter_output}
\end{align}
where $Z_m(s)=\mathcal{L}\{z_m(t)\}$.

Define the filter state $\mathbf{x}_m$ as: 
\begin{align}
    \mathbf{x}_m
\coloneqq
\begin{bmatrix}
q_m &
\dot q_m
\end{bmatrix}^{\mathsf T}\in\R^2.
\end{align}
The state-space dynamic model of the filter is formulated as:
\begin{subequations} \label{eq:continu}
\begin{align}
\dot{\mathbf{x}}_m
&=
A_m^c\mathbf{x}_m
+
B_m^c P_{\mathrm{g}},
\label{eq:mode_filter_continuous_state}
\\
z_m
&=
C_m^c\mathbf{x}_m,
\label{eq:mode_filter_continuous_output}
\end{align}
\end{subequations}
where the matrices are defined as:
\begin{subequations}\label{eq:mode_filter_continuous_matrices1}
\begin{align}
A_m^c
&=
\begin{bmatrix}
0 & 1\\
-\omega_m^2 & -2\zeta_{m}\omega_m
\end{bmatrix},
\quad
B_m^c
=
\begin{bmatrix}
0\\
1
\end{bmatrix},\\
C_m^c
&=
\begin{bmatrix}
0 &
2\zeta_{m}\omega_m
\end{bmatrix}.
\end{align}
\end{subequations}

To facilitate the implementation of discrete-time MPC, the continuous-time dynamics \eqref{eq:continu} are  discretized with a sampling interval $\Delta t$, and the resulting discrete-time mode extraction dynamics are given by \eqref{eq:discrete}:
\begin{subequations}\label{eq:discrete}
    \begin{align}
\mathbf{x}_{m,k+1}
&=
A_m\mathbf{x}_{m,k}
+
B_mP_{\mathrm{g},k},
\label{eq:mode_filter_discrete_state}
\\
z_{m,k}
&=
C_m\mathbf{x}_{m,k},
\label{eq:mode_filter_discrete_output}
\end{align}
\end{subequations}
where $k$ denotes the discrete time step, and
\begin{align*}
A_m
=
e^{A_m^{c}\Delta t},\ 
B_m
=
\int_{0}^{\Delta t}\!
e^{A_m^{c}\tau}
B_m^{c}
\,d\tau,\ 
C_m
=
C_m^{c}.
\end{align*}
The variable $z_{m,k}$ denotes the component of the grid-side data center power within the frequency band surrounding the vulnerable mode $\omega_m$. By penalizing or constraining $z_{m,k}$, the proposed controller suppresses the data center power variations that can excite the corresponding poorly damped grid mode.

\subsection{Control Design Problem}

In this paper, we consider a two-level control architecture for the HESS: the higher-level controller sets the power references to coordinate the BESS and SC, while the lower-level converter controllers closely track these power references. This architecture is consistent with practical converter-based energy storage systems in data centers \cite{Wang2026AIDCPSCAD}, where the lower-level converter controllers typically operate in a fast timescale of milliseconds to tens of milliseconds, while the higher-level controller operates in a slower timescale of hundreds of milliseconds or seconds. 
To capture the fast power tracking dynamics of the power converters in the HESS, the BESS and SC power references generated by the higher-level controller are passed through the approximate first-order dynamic model \eqref{eq:storage_power_dynamics} to obtain their actual power outputs:
\begin{subequations}\label{eq:storage_power_dynamics}
   \begin{align}
    \tau_{\tb} \dot{P}_{\tb}^{\mathrm{act}}(t)
  &  =
    -P_{\tb}^{\mathrm{act}}(t)+P_{\tb}^{\mathrm{ref}}(t), \\
    \tau_{\ts} \dot{P}_{\ts}^{\mathrm{act}}(t)
   & =
    -P_{\ts}^{\mathrm{act}}(t)+P_{\ts}^{\mathrm{ref}}(t),
\end{align} 
\end{subequations}
where $P_{\tb/\ts}^{\mathrm{ref}}(t)$ and $P_{\tb/\ts}^{\mathrm{act}}(t)$ denote the power references and actual net power outputs of the BESS and SC at time $t$, respectively, and $\tau_{\tb/\ts}$ denotes their response time constants. 

The goal of this work is to design the higher-level controller that determines the power reference trajectories $P_{\tb/\ts}^{\mathrm{ref}}(t)$ to coordinate the BESS and SC for power smoothing of the AI data center. For ease of notation, we omit the superscript ``$\mathrm{ref}$" in $P_{\tb/\ts}^{\mathrm{ref}}$ and use $P_{\tb/\ts}$ hereafter.


\section{Grid-Mode-Aware Model Predictive Control for Hybrid Energy Storage System}
\label{sec:mpc}

In this section, the G-MPC optimization models are formulated, and a solution algorithm is developed to coordinate the BESS and SC for AI data center load smoothing.

Consider a discrete-time horizon with the sampling  interval $\Delta t$.
At each time step, the proposed G-MPC optimizes the BESS and SC power trajectories over an $N$-step prediction horizon based on the predicted AI data center demand. Only the control action corresponding to the first prediction step is implemented. Then, at the next time step, the prediction horizon is shifted forward by one step, the AI data center power prediction and system states are updated, and the G-MPC optimization problem is solved again. This receding-horizon procedure enables the HESS to continuously adapt its control actions to evolving data center power fluctuations and system operating conditions. For ease of notation, an instance of the G-MPC optimization problem is presented below over an $N$-step prediction horizon, denoted by $[N]\coloneqq\{1,2,\cdots,N\}$.

\subsection{Energy Storage Dynamics and Constraints}
\label{subsec:storage_constraints}


Let $E_{\tb,k}$ and $E_{\ts,k}$ denote the stored energy of the BESS and SC at time step $k$, and their discrete-time dynamics are given by \eqref{eq:energy_dynamics} for all $k\in[N]$:
\begin{subequations} \label{eq:energy_dynamics}
    \begin{align}
E_{\tb,k}
&=  E_{\tb,k-1} + \Delta t \Big(\eta_{\tb}^{\mathrm{ch}}
P_{\tb,k}^{\mathrm{ch}}
-
\frac{1}{\eta_{\tb}^{\mathrm{dis}}}P_{\tb,k}^{\mathrm{dis}}
\Big),
\label{eq:bat_energy_dynamics}\\
E_{\ts,k}
&=  E_{\ts,k-1} + \Delta t \Big(\eta_{\ts}^{\mathrm{ch}}
P_{\ts,k}^{\mathrm{ch}}
-
\frac{1}{\eta_{\ts}^{\mathrm{dis}}}P_{\ts,k}^{\mathrm{dis}}
\Big),
\label{eq:sc_energy_dynamics}
\end{align}
\end{subequations}
where $\eta_{\tb}^{\mathrm{ch}},\eta_{\tb}^{\mathrm{dis}}\in (0,1]$ and $\eta_{\ts}^{\mathrm{ch}},\eta_{\ts}^{\mathrm{dis}}\in (0,1]$ denote the charging and discharging efficiencies of the BESS and SC, respectively, while $E_{\tb,0}$ and $E_{\ts,0}$ denote the initial stored energy of the BESS and SC. Here,
$P_{\tb,k}^{\mathrm{ch}}, P_{\tb,k}^{\mathrm{dis}}$ and $P_{\ts,k}^{\mathrm{ch}}, P_{\ts,k}^{\mathrm{dis}}$
denote the charging and discharging power of the BESS and SC at time step $k$, respectively; and their net power outputs are given by \eqref{eq:powerout}:
\begin{align}\label{eq:powerout}
    P_{\tb,k} =  P_{\tb,k}^{\mathrm{dis}} - P_{\tb,k}^{\mathrm{ch}}, \   P_{\ts,k} =  P_{\ts,k}^{\mathrm{dis}} - P_{\ts,k}^{\mathrm{ch}}.
 \end{align}

The corresponding state-of-charge (SOC) constraints of the BESS and SC are given by \eqref{eq:storage_energy_limits} for all $k\in [N]$:
\begin{subequations}\label{eq:storage_energy_limits}
\begin{gather}
 \underline E_\tb
\leq
 E_{\tb,k}
\leq
\overline E_\tb,
\\
\underline E_\ts
\leq E_{\ts,k}
\leq
\overline E_\ts,
\end{gather}
\end{subequations}
where $\underline E_\tb, \overline E_\tb$ and $\underline E_\ts, \overline E_\ts$ denote the lower and upper energy limits of the BESS and SC, respectively. 
Typically, the SC has a relatively small energy capacity due to its low energy density and high capital cost  \cite{durvasulu2023technology}, while the BESS
has a larger energy capacity to 
compensate for sustained power variations,  with $\overline{E}_{\tb} \gg \overline{E}_{\ts}$.

The charging and discharging powers of the BESS and SC satisfy the constraints \eqref{eq:powercon} for all $k\in[N]$:
\begin{subequations} \label{eq:powercon}
    \begin{gather} 
0
\!\leq\!
P_{\tb,k}^{\mathrm{ch}}
\!\leq\! \alpha_{k}^\tb
\overline  P_\tb^{\mathrm{ch}}, \ 0
\!\leq\!
P_{\tb,k}^{\mathrm{dis}}
\!\leq\! (1\!-\!\alpha_{k}^\tb)
\overline P_\tb^{\mathrm{dis}}, \label{eq:powercon:bat}
\\
0
\!\leq\!
P_{\ts,k}^{\mathrm{ch}}
\!\leq\! \alpha_{k}^\ts
\overline P_\ts^{\mathrm{ch}}, \ 0
\!\leq\!
P_{\ts,k}^{\mathrm{dis}}
\!\leq\! (1-\alpha_{k}^\ts)
\overline P_\ts^{\mathrm{dis}}, \label{eq:powercon:sc} 
\end{gather}
\end{subequations}
which represent the charging and discharging power limits with the  parameters $\overline P_\tb^{\mathrm{ch}},\overline P_\tb^{\mathrm{dis}},\overline P_\ts^{\mathrm{ch}},\overline P_\ts^{\mathrm{dis}}$. The binary variables $\alpha_{k}^\tb$ and $\alpha_{k}^\ts$ are introduced to prevent simultaneous charging and discharging, which satisfy \eqref{eq:binary} for all $k\in[N]$:
\begin{align}\label{eq:binary}
    \alpha_{k}^\tb \in \{0,1\}, \ \alpha_{k}^\ts \in \{0,1\}.
\end{align}
A key issue regarding these binary constraints is that they render the optimization problem nonconvex, substantially increasing the computational complexity. To address this issue, a tailored solution algorithm is developed in Section~\ref{subsec:complete_qp}.


\subsection{Grid-Side Power Requirements}
\label{subsec:grid_constraints}

Given the AI data center power demand $P_{\mathrm{dc},k}$ at each time step $k\in[N]$, the grid-side load power is $P_{\mathrm{g},k}$: 
\begin{equation}
P_{\mathrm{g},k} = P_{\mathrm{dc},k}
-
P_{\mathrm{bat},k}
-
P_{\mathrm{sc},k}.
\label{eq:grid_power_balance}
\end{equation} 
To accommodate occasional large AI load transitions without rendering
the G-MPC problem infeasible, the grid-side power envelope \eqref{eq:grid_power_envelope} and ramp rate requirement \eqref{eq:grid_ramp_setup} are implemented as the soft constraints \eqref{eq:enve:soft} and \eqref{eq:ramp:soft}, respectively.
\begin{gather} 
  \underline{P}_{\mathrm{g}} - \epsilon_{p,k}\leq P_{\mathrm{g},k} \leq \overline{P}_{\mathrm{g}} + \epsilon_{p,k}, \label{eq:enve:soft}\\
  -R_{\mathrm{g}}^{\mathrm{down}} -  \epsilon_{r,k}
\leq
\frac{1}{\Delta t}(P_{\mathrm{g},k} - P_{\mathrm{g},k-1})
\leq
R_{\mathrm{g}}^{\mathrm{up}} + \epsilon_{r,k},\label{eq:ramp:soft}\\
\epsilon_{p,k}\geq 0,\ \ \epsilon_{r,k}\geq 0.
\end{gather}
Here, nonnegative slack variables $\epsilon_{p,k}, \epsilon_{r,k}$ are introduced at each time step $k\in[N]$ to relax the power envelope and ramp rate requirements. These slack variables will be penalized in the objective function to discourage constraint violations while preserving the feasibility of the optimization problem. 

Moreover, to limit the grid-side power components around vulnerable grid oscillatory modes, the following soft constraint \eqref{eq:modecon} is imposed for each $m\in\mathcal{M}$ and $k\in[N]$:
\begin{align} \label{eq:modecon}
    -\overline z_m-\epsilon_{m,k} \leq z_{m,k}
\leq \overline z_m+\epsilon_{m,k},\ \ \epsilon_{m,k}\geq 0,
\end{align}
where $\overline z_m$ is the prescribed upper bound on the magnitude of the grid-side power component around vulnerable mode $m$, as specified by the grid operator. 
The nonnegative slack variable $\epsilon_{m,k}$ relaxes this modal-power limit when necessary and will be penalized in the objective function to discourage violations while maintaining the feasibility of the optimization problem.

\subsection{Definition of G-MPC Objective}
\label{subsec:objective}

The proposed G-MPC incorporates both the grid-side and HESS-side objectives, as elaborated below.

\subsubsection{Grid-side Objective}
The grid-side objective function at each time step $k\in[N]$ is defined as \eqref{eq:grid:obj}:
\begin{align}\label{eq:grid:obj}
    J_{\mathrm{grid},k}=   c_{p,k}\epsilon_{p,k}^2 +  c_{r,k}\epsilon_{r,k}^2 + \sum_{m\in\mathcal{M}}\!c_{m,k}\epsilon_{m,k}^2, 
\end{align}
where the three terms penalize the slack variables associated with the violations of the power envelope, ramp-rate, and grid-mode suppression requirements, respectively.
The corresponding cost coefficients are $c_{p,k}$, $c_{r,k}$, and $c_{m,k}$, while quadratic penalties are adopted so that larger constraint violations incur increasingly higher marginal costs. 

\subsubsection{HESS-side Objective} The HESS operating cost is primarily associated with the degradation of the BESS and SC due to charging and discharging. A widely-used method models the degradation cost as proportional to the energy throughput \cite{liu2018optimal,bera2020maximising}, given by \eqref{eq:hess:deg} for time step $k\in[N]$:
\begin{align}\label{eq:hess:deg}
J_{\mathrm{hess},k}^{\mathrm{deg}}
&=  c_{\tb}^{\mathrm{deg}}\cdot
\Big(
 \eta_{\tb}^{\mathrm{ch}}
P_{\tb,k}^{\mathrm{ch}}
+
\frac{1}{\eta_{\tb}^{\mathrm{dis}}}P_{\tb,k}^{\mathrm{dis}}
\Big) \nonumber \\
&\qquad 
+ c_{\ts}^{\mathrm{deg}}\cdot
\Big(\eta_{\ts}^{\mathrm{ch}}
P_{\ts,k}^{\mathrm{ch}}
+
\frac{1}{\eta_{\ts}^{\mathrm{dis}}}P_{\ts,k}^{\mathrm{dis}}
\Big),
\end{align}
where $c_{\tb}^{\mathrm{deg}}$ and $c_{\ts}^{\mathrm{deg}}$ denote the marginal degradation costs per unit of energy throughput of the BESS and SC, respectively. Since SC typically offers an extremely long cycle life, on the order of millions of cycles \cite{durvasulu2023technology}, its degradation cost coefficient is substantially lower than the BESS, i.e., $c_{\tb}^{\mathrm{deg}} \gg c_{\ts}^{\mathrm{deg}}$.

In addition, power-ramping penalties are introduced for both the BESS and SC, as given by \eqref{eq:bess:ramp:obj}:
\begin{align}\label{eq:bess:ramp:obj}
    J_{\mathrm{hess},k}^{\mathrm{ramp}}= \, &c_\tb^{\Delta}\cdot  \Big(\frac{ P_{\tb,k}
- P_{\tb,k-1}}{\Delta t}\Big)^2  \nonumber\\ 
&\qquad  + c_\ts^{\Delta}\cdot  \Big( \frac{P_{\ts,k}
- P_{\ts,k-1}}{\Delta t}\Big)^2, 
\end{align}
where $c_{\tb}^{\Delta}$ and $c_{\ts}^{\Delta}$ denote the corresponding ramping penalty coefficients with $c_{\tb}^{\Delta}\gg c_{\ts}^{\Delta}$. A larger penalty on rapid BESS power variations discourages the battery from compensating for fast load fluctuations and allows the SC to absorb a greater share of them. This achieves the desired BESS-SC power allocation without decomposing the data center load variations by a predefined cut-off frequency.
Accordingly, the total HESS operating cost at time step $k\in[N]$ is given by \eqref{eq:hess:cost}: 
\begin{align}\label{eq:hess:cost}
    J_{\mathrm{hess},k} = J_{\mathrm{hess},k}^{\mathrm{deg}}+J_{\mathrm{hess},k}^{\mathrm{ramp}}.
\end{align}

\subsection{Complete G-MPC Models and Solution Algorithm}
\label{subsec:complete_qp}

The complete G-MPC optimization model is formulated as: 
\begin{subequations}\label{eq:exact}
    \begin{align}
     J^{\star}_{\mathrm{exact}}=   \min & \sum_{k\in[N]}\!\!\Delta t\big(J_{\mathrm{grid},k} + J_{\mathrm{hess},k}\big),\\
        \mathrm{s.t.}\ & \eqref{eq:discrete}, \eqref{eq:energy_dynamics}-\eqref{eq:hess:cost}.
    \end{align}
\end{subequations}
The resulting model \eqref{eq:exact} is a mixed-integer quadratic programming (MIQP) problem due to the binary constraints \eqref{eq:binary}. The  nonconvexity and computational burden make it challenging to solve efficiently for real-time HESS control. 

To address this challenge, we introduce a relaxed optimization model \eqref{eq:relax} to enable efficient solution: 
\begin{subequations}\label{eq:relax}
    \begin{align}
     J^{\star}_{\mathrm{relax}}=   \min & \sum_{k\in[N]}\!\!\Delta t\big(J_{\mathrm{grid},k} + J_{\mathrm{hess},k}\big),\\
        \mathrm{s.t.}\ & \eqref{eq:discrete}, \eqref{eq:energy_dynamics}-\eqref{eq:powercon}, \eqref{eq:grid_power_balance}-\eqref{eq:hess:cost},\\
        & \alpha_{k}^\tb \in [0,1], \ \alpha_{k}^\ts \in [0,1]. \label{eq:relax:bin}
    \end{align}
\end{subequations}
which replaces the binary constraints \eqref{eq:binary} with their continuous relaxation \eqref{eq:relax:bin}. The relaxed model \eqref{eq:relax} is a convex quadratic programming (QP) problem with linear constraints and can be efficiently solved using standard QP solvers.
Nevertheless, the relaxed model \eqref{eq:relax} may yield optimal solutions with simultaneous charging and discharging.

To this end,
a \emph{fix-and-re-optimize} algorithm (Algorithm~\ref{alg:reopt}) is introduced to obtain solutions that ensure no simultaneous charging and discharging for the BESS and SC. Specifically, after solving the relaxed model \eqref{eq:relax}, if simultaneous charging and discharging appears in the optimal solution, the operating status of the BESS and SC is recovered from their net energy changes and net power outputs. For example, if
$\eta_{\tb}^{\mathrm{ch}}
P_{\tb,k}^{\mathrm{ch}}
-
\frac{1}{\eta_{\tb}^{\mathrm{dis}}}P_{\tb,k}^{\mathrm{dis}}$ is positive, the BESS is identified as operating in the charging status at time step $k$. We then fix $\alpha_k^\tb \!=\! 1$, together with all other binary variables determined in the same manner,
and re-solve the resulting QP to obtain the final solution. Algorithm \ref{alg:reopt} presents the detailed procedure.
In the classification rule \eqref{eq:fixrule}, small tolerance values $\varepsilon_E, \varepsilon_P$ are introduced to avoid numerical ambiguity when the net stored-energy change or net power is close to zero.

\begin{algorithm}
\caption{Fix-and-Re-Optimize Algorithm for G-MPC}
\label{alg:reopt}
\begin{algorithmic}[1]

\State Update the predicted data center load and initialize system states using the solution from the previous G-MPC step.

\State Solve the relaxed QP model \eqref{eq:relax}.

\If{The complementary condition ($P^{\mathrm{ch}}_k P^{\mathrm{dis}}_k =0$) is violated at any time step $k\in[N]$ for the BESS or SC}

 \State For all $k\in[N]$, compute the predicted stored energy change for the BESS and SC:
        \[
        \Delta E^{\tb/\ts}_k
        =
        \eta_{\tb/\ts}^{\mathrm{ch}}P_{\tb/\ts,k}^{\mathrm{ch}}
        -
        \frac{1}{\eta_{\tb/\ts}^{\mathrm{dis}}}
        P_{\tb/\ts,k}^{\mathrm{dis}}.
        \]
\State Fix $\alpha_k^{\tb/\ts}$ according to the classification rule \eqref{eq:fixrule}: 
\begin{equation}\label{eq:fixrule}
\alpha_k^{\tb/\ts}
=
\begin{cases}
1, & \Delta E^{\tb/\ts}_k \!>\! \varepsilon_E,\\
0, & \Delta E^{\tb/\ts}_k \!<\! -\varepsilon_E,\\
0, & |\Delta E^{\tb/\ts}_k| \!\le\! \varepsilon_E,\; P_{\tb/\ts,k} \!>\! \varepsilon_P,\\
1, & |\Delta E^{\tb/\ts}_k| \!\le\! \varepsilon_E,\; P_{\tb/\ts,k} \!<\! -\varepsilon_P.
\end{cases}
\end{equation}

If both $|\Delta E^{\tb/\ts}_k| \!\le\! \varepsilon_E$ and $|P_{\tb/\ts,k}| \!\le\! \varepsilon_P$, fix
\begin{align*}
    P_{\tb/\ts,k}^{\mathrm{ch}} = P_{\tb/\ts,k}^{\mathrm{dis}} = 0.
\end{align*}
        \State Solve the QP model \eqref{eq:fix} with the fixed values: 
\begin{subequations}\label{eq:fix}
    \begin{align}
     J^{\star}_{\mathrm{fix}}=   \min & \sum_{k\in[N]}\!\!\Delta t\big(J_{\mathrm{grid},k} + J_{\mathrm{hess},k}\big),\\
        \mathrm{s.t.}\ & \eqref{eq:discrete}, \eqref{eq:energy_dynamics}-\eqref{eq:powercon}, \eqref{eq:grid_power_balance}-\eqref{eq:hess:cost}.
    \end{align}
\end{subequations}
\EndIf

\State Implement the first-step optimal BESS and SC control actions $(P_{\tb,1}^{\mathrm{ch}\star},P_{\tb,1}^{\mathrm{dis}\star},P_{\ts,1}^{\mathrm{ch}\star}, P_{\ts,1}^{\mathrm{dis}\star})$.
\State Shift the optimization time horizon one-step forward and initialize the next G-MPC step.

\end{algorithmic}
\end{algorithm}

For Algorithm~\ref{alg:reopt}, at most two QP problems need to be solved at each MPC step: the relaxed QP \eqref{eq:relax} is solved first; and only when simultaneous charging and discharging is detected, a second QP \eqref{eq:fix} with fixed status is solved after determining the corresponding charging/discharging status. Therefore, the final HESS control decisions avoid simultaneous charging and discharging, while retaining the computational efficiency of QP-based optimization. 
Since the grid-side power requirements are imposed as soft constraints, all three optimization problems \eqref{eq:exact}, \eqref{eq:relax}, \eqref{eq:fix} remain feasible.
By construction, the relaxed problem \eqref{eq:relax} provides a lower bound on the objective of the original problem \eqref{eq:exact}, whereas the solution of the fixed-status problem \eqref{eq:fix} must be feasible for the original problem \eqref{eq:exact}. Hence, their optimal objective values satisfy:
\begin{align}
    J^{\star}_{\mathrm{relax}} \leq J^{\star}_{\mathrm{exact}}\leq J^{\star}_{\mathrm{fix}}.
\end{align}
If the fix-and-re-optimize step in Algorithm \ref{alg:reopt} is not activated (i.e., no simultaneous charging and discharging), the optimal solution of the relaxed model \eqref{eq:relax} is also optimal for the original model \eqref{eq:exact} with $J^{\star}_{\mathrm{relax}} \!=\! J^{\star}_{\mathrm{exact}}$. If it is activated,
the objective value difference $J^{\star}_{\mathrm{fix}}\!-\!J^{\star}_{\mathrm{relax}} $ provides a computable upper bound on the optimality gap of the final solution.



\section{Numerical Simulations}\label{sec:simulation}

In this section, extensive simulations are conducted to evaluate the performance and characteristics of the proposed G-MPC-based HESS control algorithm for AI data center power smoothing and grid oscillation mitigation.

\subsection{Simulation Setup and Parameter Settings}
\label{subsec:case_setup}




\subsubsection{Aggregate Power Grid Dynamic Model}
The AI data center is connected to a power system represented by an equivalent synchronous generator to evaluate the impact of its power variations on grid frequency. The aggregate grid frequency dynamics are modeled using the widely-used linearized swing and governor equations \eqref{eq:aggregate_grid}:
\begin{subequations} \label{eq:aggregate_grid}
\begin{align}
   M_{\mathrm{g}}\dot{\Delta f}(t)
    &=
    \Delta P_{\mathrm{m}}(t)
    -
    \frac{\Delta P_{\mathrm{g}}(t)}{P_\mathrm{base}}
    -
    D_{\mathrm{g}}\Delta f(t),\label{eq:grid:swing} \\
    T_{\mathrm{gov}}\dot{\Delta P}_{\mathrm{m}}(t)
    &=
    -\Delta P_{\mathrm{m}}(t)
    -
    \frac{1}{R_{\mathrm{gov}}}\Delta f(t), \label{eq:grid:gen}
\end{align}
\end{subequations}
Here, $\Delta f(t)$ denotes the deviation of the grid frequency 
from its nominal value at time $t$, $\Delta P_{\mathrm{m}}(t)$ denotes the 
deviation in the mechanical power supplied by the aggregate synchronous 
generator, and $\Delta P_{\mathrm{g}}(t)
= P_{\mathrm{g}}(t)-P_{\mathrm{g}}^{\mathrm{ave}}$ denotes the deviation of the grid-side data center load from its average value. The parameters $M_{\mathrm{g}}$ and 
$D_{\mathrm{g}}$ represent the aggregate grid inertia and damping 
coefficients, respectively, while $T_{\mathrm{gov}}$ and 
$R_{\mathrm{gov}}$ denote the governor time constant and droop coefficient 
of the generator, respectively. $P_\mathrm{base}$ is the base power value for scaling. Equation~\eqref{eq:grid:swing} 
describes the grid frequency response to instantaneous power imbalance, 
while \eqref{eq:grid:gen} represents the primary frequency regulation 
provided by the governor of the synchronous generator. Accordingly, the natural angular frequency $\omega_{\mathrm{nat}}$ of the grid dynamics \eqref{eq:aggregate_grid} is given by \eqref{eq:natural}:
\begin{align}\label{eq:natural}
\omega_{\mathrm{nat}} = 2\pi f_{\mathrm{nat}}= \sqrt{\frac{D_\mathrm{g} +1/R_{\mathrm{gov}}}{M_{\mathrm{g}}T_{\mathrm{gov}}}}.
\end{align}
As only one oscillatory mode is present, we define $\mathcal{M}\coloneqq\{1\}$ and use $\omega_1=\omega_{\mathrm{nat}}$ as the targeted modal angular frequency in \eqref{eq:modal_bandpass_filter} and \eqref{eq:mode_filter_continuous_matrices1}. Given the system parameters in Table~\ref{tab:parameters}, the corresponding natural frequency is $f_{\mathrm{nat}}\!=\!\frac{\omega_{\mathrm{nat}}}{2\pi}\!=\!0.5$~Hz. The damping ratio of the modal filter in \eqref{eq:modal_bandpass_filter} is set to $\zeta_1=0.1$.


\subsubsection{AI Data Center Load Profile and HESS Configuration}

The real measured GPU power demand profile during AI model training in \cite{choukse2025power} is scaled to a peak of 100~MW and combined with a constant 50~MW load representing cooling and other auxiliary systems. Moreover, a periodic component of $20\sin(\omega_{\mathrm{nat}}t)$~MW at the natural angular frequency $\omega_{\mathrm{nat}}$ of the aggregate grid is further added to construct the time-varying data center load $P_{\mathrm{dc}}(t)$, as shown in Figure~\ref{fig:power}. The HESS consists of a $40$-MW/$40$-MWh BESS and a $20$-MW/$0.3$-MWh SC, both of which operate within a $10\%$-$90\%$ energy window. 
Their initial stored energy values are selected as $E_{\tb,0}\!=\!(\underline E_\tb \!+\! \overline E_\tb)/{2}$ and $E_{\ts,0}\!=\!(\underline E_\ts \!+\! \overline E_\ts)/{2}$
 to provide symmetric charging and discharging capabilities. 
At each control step $[t, t+\Delta t]$, the actual power outputs of the BESS and SC are obtained by applying the power references generated by the G-MPC method to the converter reference-tracking dynamics in \eqref{eq:storage_power_dynamics}. 
 
 For the G-MPC implementation, the sampling interval is $\Delta t=0.25$s, and a $N=200$ prediction horizon is used, corresponding to a $50$s prediction window. The total simulation time horizon is $300$s. 
 The main system parameters and cost coefficients are summarized in Table \ref{tab:parameters}.

\begin{table}[t]
\centering
\caption{System Parameters and Cost Coefficients.}
\label{tab:parameters}
\renewcommand{\arraystretch}{1.05}
\setlength{\tabcolsep}{2.2pt}
\scriptsize
\resizebox{\columnwidth}{!}{%
\begin{tabular}{@{}cccccc@{}}
\cline{1-2}\cline{3-4}\cline{5-6}
\textbf{Notation} & \textbf{Value}
& \textbf{Notation} & \textbf{Value}
& \textbf{Notation} & \textbf{Value}\\
\hline

$c_p$
& 10~\$/{MW}$^{2}$h
& $\underline{E}_{\mathrm{bat}}$
& 4~MWh
& $\underline{E}_{\mathrm{sc}}$
& 0.03~MWh\\

$c_r$
& 3~\$/(MW/s)$^{2}$h
& $\overline{E}_{\mathrm{bat}}$
& 36~MWh
& $\overline{E}_{\mathrm{sc}}$
& 0.27~MWh\\

$c_m$
& 500~\$/{MW}$^{2}$h
& $E_{\mathrm{bat},0}$
& 20~MWh
& $E_{\mathrm{sc},0}$
& 0.15~MWh\\

$c_{\mathrm{bat}}^{\mathrm{deg}}$
& 20~\$/{MWh}
& $\overline{P}_{\mathrm{bat}}^{\mathrm{ch/dis}}$
& 40~MW
& $\overline{P}_{\mathrm{sc}}^{\mathrm{ch/dis}}$
& 20~MW\\

$c_{\mathrm{sc}}^{\mathrm{deg}}$
& 5~\$/{MWh}
& $\eta_{\mathrm{bat}}^{\mathrm{ch/dis}}$
& 0.97
& $\eta_{\mathrm{sc}}^{\mathrm{ch/dis}}$
& 0.98\\

$c_{\mathrm{bat}}^{\Delta}$
& 2~\$/({MW/s})$^{2}$h
& $\tau_{\mathrm{bat}}$
& 0.08~s
& $\tau_{\mathrm{sc}}$
& 0.05~s\\

$c_{\mathrm{sc}}^{\Delta}$
& 0.2~\$/({MW/s})$^{2}$h
&$\Delta t$
& 0.25~s & $N$& 200\\

$M_\mathrm{g}$& 0.00859~s$^2$
& $R_{\mathrm{g}}^{\mathrm{down}}$& 1~MW/s & 
$\overline{P}_{\mathrm{g}} $
& 130~MW\\

$D_\mathrm{g}$
& 0.00595~s
& $R_{\mathrm{g}}^{\mathrm{up}}$&  1~MW/s & $\underline{P}_{\mathrm{g}} $
& 120~MW\\

$R_{\mathrm{gov}}$& 3.0~s$^{-1}$
& $T_{\mathrm{gov}}$&  4~s &$P_{\mathrm{base}}$ & $1\!\times\! 10^4$~MW 
 \\

 $\zeta_{1}$& 0.1
& $\omega_1$& 3.141 rad/s & $\overline z_1$& 0~MW
 \\

\hline
\end{tabular}%
}
\end{table}

\vspace{3pt}
To evaluate the effectiveness of the proposed G-MPC method in suppressing the targeted grid oscillatory modes and mitigating grid impacts, we compare the following three cases:
\begin{itemize}

\item \emph{Case 1 (Complete Model):} Both the targeted mode suppression objective and the grid-side power envelope and ramp-rate requirements are considered in \eqref{eq:grid:obj}. This corresponds to the complete model in Section \ref{subsec:complete_qp}.

\item \emph{Case 2 (Mode Suppression Only):} Only the targeted grid mode suppression objective is considered in \eqref{eq:grid:obj}, while the grid-side power envelope and ramp-rate requirements are excluded by setting $c_{p,k}=c_{r,k}=0$.

\item \emph{Case 3 (Without Explicit Mode Suppression):} Only the grid-side power envelope and ramp-rate requirements are considered in \eqref{eq:grid:obj}, while the targeted grid mode suppression objective is excluded by setting $c_{m,k}=0$.

\end{itemize}

In the following, Section \ref{subsec:complete} presents the simulation results for the complete model in Case 1, demonstrating the performance and characteristics of the proposed G-MPC method. Section \ref{subsec:compare} then compares the three cases above in terms of targeted mode suppression and the resulting grid frequency responses. Section \ref{subsec:SOCenforce} presents the results 
under additional terminal conditions that restore the SOC levels of BESS and SC to their initial values. Finally, the computational efficiency of the proposed method is provided in \ref{subsec:efficiency}.

\subsection{Performance of the Proposed G-MPC Framework}\label{subsec:complete}

The proposed G-MPC-based HESS control algorithm (Algorithm \ref{alg:reopt}) is implemented to coordinate the operation of the BESS and SC to smooth AI data center power variations. The simulation results in Case 1 are presented below.

\subsubsection{Data Center Load and HESS Power Outputs}
\begin{figure}
    \centering
    \includegraphics[width=1\linewidth]{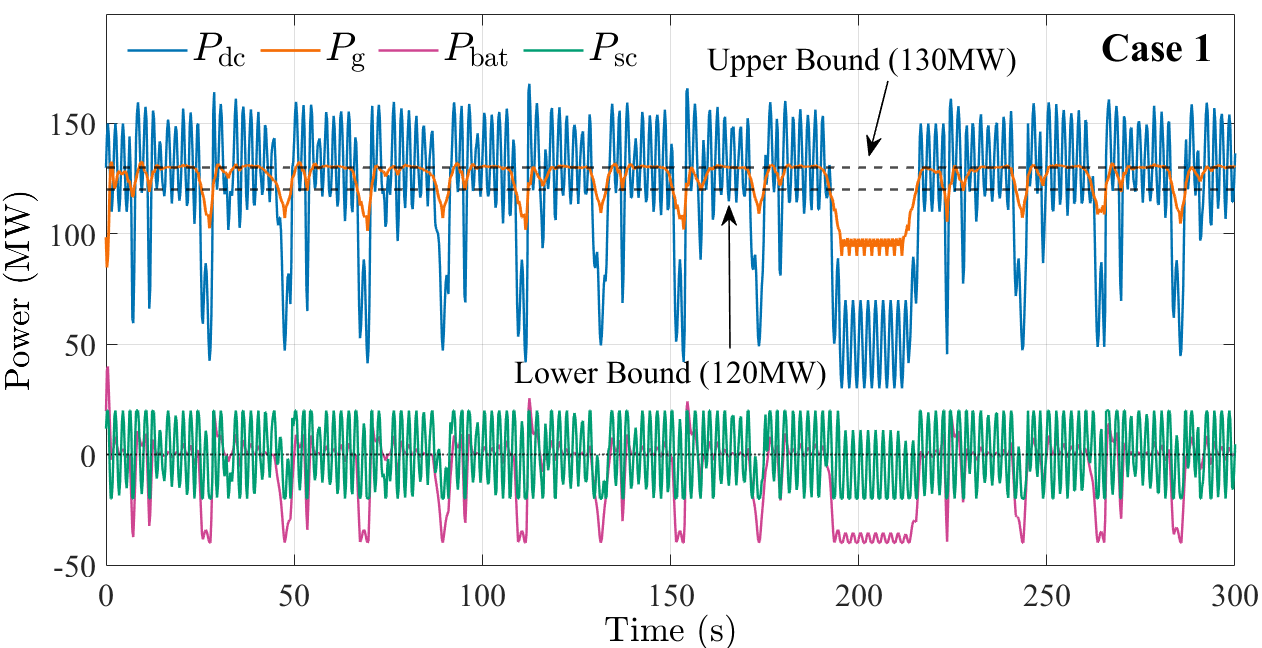}
    \caption{AI data center power demand $P_{\mathrm{dc}}(t)$, grid-side load $P_{\mathrm{g}}(t)$, BESS and SC power outputs $P_\tb(t),P_\ts(t)$ under the proposed G-MPC method in Case 1.}
    \label{fig:power}
\end{figure}

Figure~\ref{fig:power} shows the AI data center power demand $P_{\mathrm{dc}}$, the grid-side load $P_{\mathrm{g}}$, and the net power output ($P_\tb,P_\ts$) of the BESS and SC. 
As shown in Figure~\ref{fig:power}, the proposed G-MPC method effectively smooths the highly fluctuating AI data center demand $P_{\mathrm{dc}}$ and maintains the grid-side load $P_{\mathrm{g}}$ within the acceptable interval of $[120,130]$~MW for most of the simulation period. Temporary violations occur only during very large demand excursions, when both the BESS and SC reach their charging or discharging power limits. Compared with the raw data center demand $P_{\mathrm{dc}}$, the grid-side load $P_{\mathrm{g}}$ exhibits substantially smaller and smoother power variations, indicating that the proposed control effectively reduces the grid-side load ramping rate. Moreover, the HESS power outputs exhibit the complementary roles of the two storage technologies: the SC responds rapidly to high-frequency demand variations, while the BESS compensates for slower and more sustained power changes. This
coordination enables the HESS to optimally leverage both the fast response of the SC and the larger energy capacity of the BESS for power smoothing.

\subsubsection{HESS Energy Trajectories and Re-Optimization}
\begin{figure}
    \centering
    \includegraphics[width=1\linewidth]{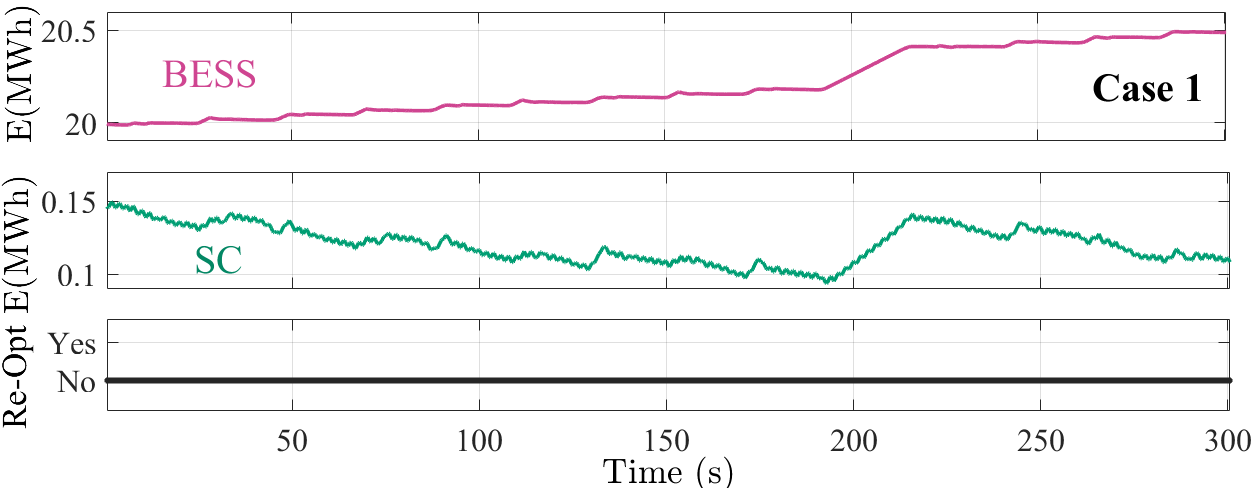}
    \caption{Stored energy $ E_{\tb}(t), E_{\ts}(t)$ of BESS and SC, and re-optimization status under the proposed G-MPC method in Case 1.}
    \label{fig:energy}
\end{figure}

Figure~\ref{fig:energy} illustrates the corresponding stored energy ($ E_{\tb}, E_{\ts}$) trajectories of the BESS and SC, together with the re-optimization status of Algorithm~\ref{alg:reopt} in the simulation. The SC energy varies frequently in response to rapid load fluctuations, while the SOC constraints \eqref{eq:storage_energy_limits} of both BESS and SC are always satisfied. Furthermore,
the re-optimization indicator remains inactive, indicating that only the relaxed QP model \eqref{eq:relax} is solved at each MPC step, and its solution naturally satisfies the complementary conditions without simultaneous charging and discharging. Thus, the fix-and-re-optimize step is not invoked in this simulation. The results with additional terminal energy constraints are discussed in Section \ref{subsec:SOCenforce}.

\subsubsection{Objective Values}

\begin{figure}
    \centering
    \includegraphics[width=1\linewidth]{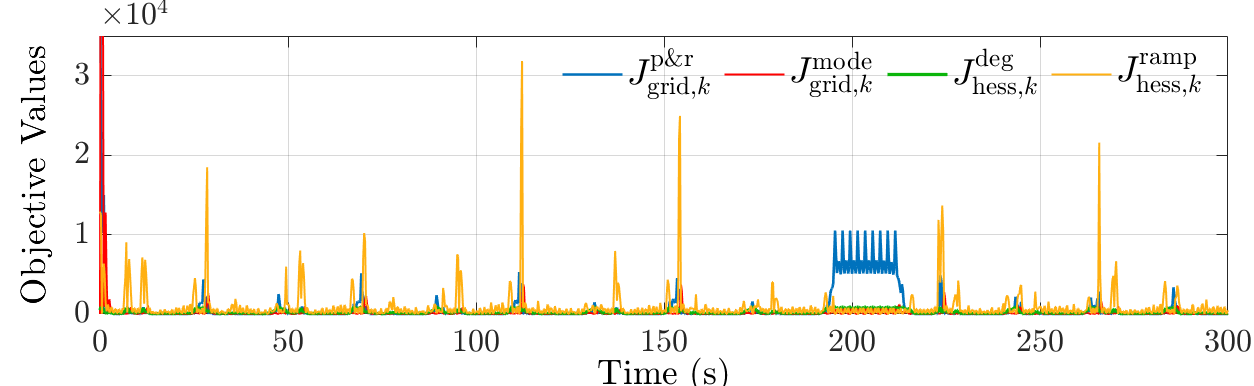}
    \caption{Objective values of the grid-side power regulation $J_{\mathrm{grid},k}^{\mathrm{p\&r}}$, targeted mode suppression $J_{\mathrm{grid},k}^{\mathrm{mode}}$, HESS degradation $ J_{\mathrm{hess},k}^{\mathrm{deg}}$, and HESS power ramping $ J_{\mathrm{hess},k}^{\mathrm{ramp}}$
    incurred at each MPC step in Case 1.}
    \label{fig:obj}
\end{figure}

Figure~\ref{fig:obj} illustrates the objective components associated with the grid-side power regulation $J_{\mathrm{grid},k}^{\mathrm{p\&r}}$, targeted mode suppression $J_{\mathrm{grid},k}^{\mathrm{mode}}$, HESS degradation $ J_{\mathrm{hess},k}^{\mathrm{deg}}$, and HESS power ramping $ J_{\mathrm{hess},k}^{\mathrm{ramp}}$ at each MPC step in Case 1. 
Specifically, $J_{\mathrm{grid},k}^{\mathrm{p\&r}}$ 
comprises the first two terms of the grid-side objective in \eqref{eq:grid:obj}, while 
$J_{\mathrm{grid},k}^{\mathrm{mode}}$ corresponds to its third term. The plotted values in Figure~\ref{fig:obj} represent the actual objective values (or operational costs) incurred by the first control action  implemented at each MPC step.
As shown in Figure~\ref{fig:obj}, the targeted mode-suppression cost $J_{\mathrm{grid},k}^{\mathrm{mode}}$ is initially large due to the initial conditions but rapidly decreases to nearly zero, demonstrating the effective suppression of the vulnerable oscillatory component. The grid-side power-regulation cost $J_{\mathrm{grid},k}^{\mathrm{p\&r}}$  remains close to zero throughout most of the simulation, except during the substantial load reduction from approximately 190s to 220s, when the power envelope requirements are temporarily violated as the BESS and SC reach their power limits. The HESS power-ramping cost $ J_{\mathrm{hess},k}^{\mathrm{ramp}}$  exhibits occasional spikes, primarily due to fast changes in the BESS power output required to accommodate rapid load variations. The HESS degradation cost $ J_{\mathrm{hess},k}^{\mathrm{deg}}$ remains relatively small throughout the simulation.

\vspace{2pt}
The next subsection investigates targeted oscillation mode suppression and its impact on the grid frequency response.




\subsection{Grid Frequency Response and Oscillation Suppression }\label{subsec:compare}

\begin{figure}
    \centering
    \includegraphics[width=1\linewidth]{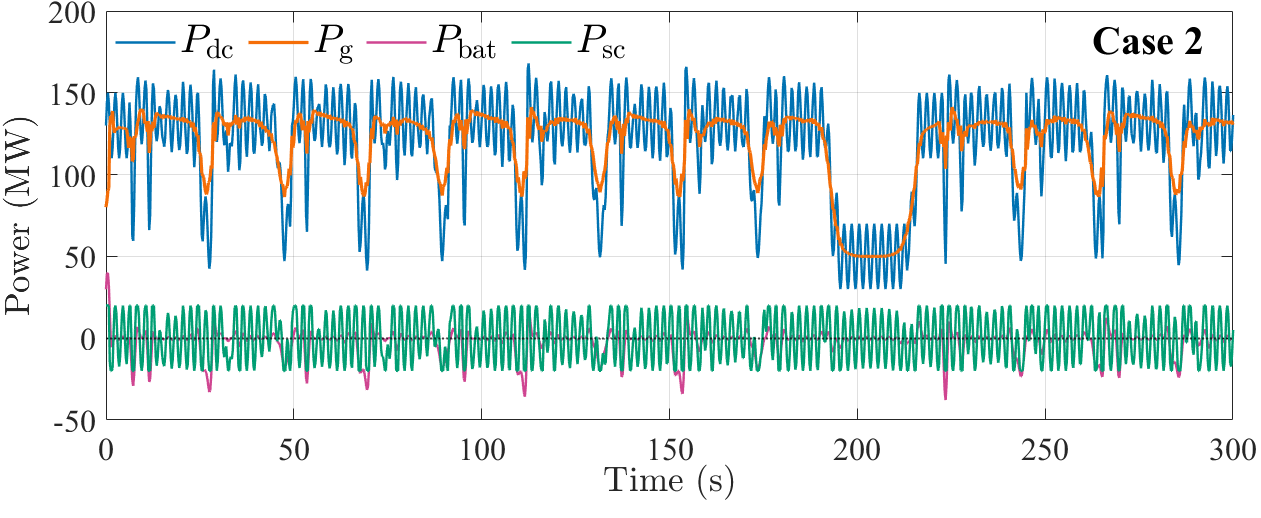}
    \includegraphics[width=1\linewidth]{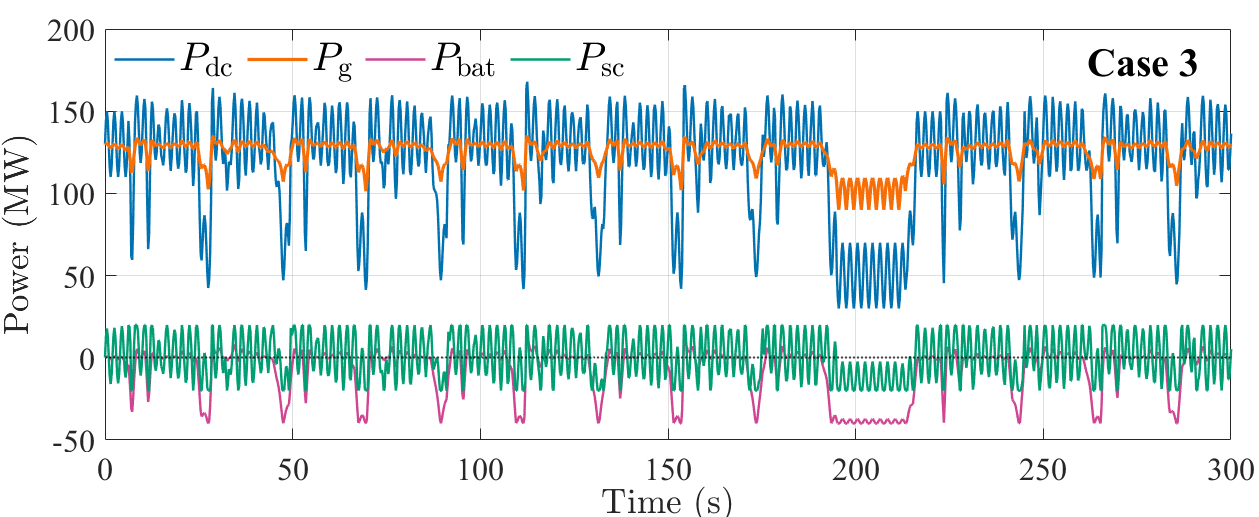}
    \caption{AI data center power demand $P_{\mathrm{dc}}(t)$, grid-side load $P_{\mathrm{g}}(t)$, BESS and SC power outputs $P_\tb(t),P_\ts(t)$ under the proposed G-MPC method in Case 2 and Case 3.}
    \label{fig:powercom}
\end{figure}

\begin{figure}
    \centering
    \includegraphics[width=1\linewidth]{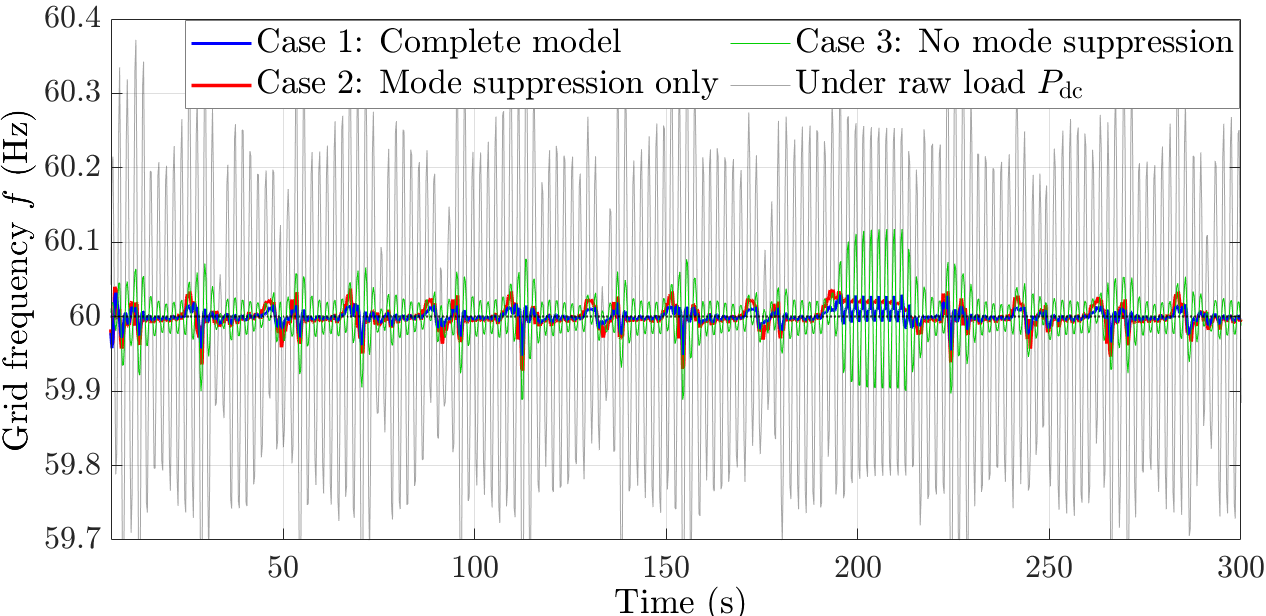}
    \caption{Comparison of the grid frequency responses under the raw data center load $P_{\mathrm{dc}}(t)$ and the HESS-filtered grid-side power $P_{\mathrm{g}}(t)$ in Cases 1-3.}
    \label{fig:frecom}
\end{figure}

The proposed G-MPC algorithm is implemented separately under Cases 1-3 for ablation studies using the same AI data center demand profile $P_{\mathrm{dc}}(t)$ introduced in Section~\ref{subsec:case_setup}. Figure~\ref{fig:powercom} shows the resulting grid-side load $P_{\mathrm{g}}(t)$ and the corresponding BESS and SC power outputs in Cases 2 and 3, while the results for Case 1 are presented in Figure~\ref{fig:power}.  
Then, the raw data center demand $P_{\mathrm{dc}}(t)$ (as a baseline) and the grid-side loads $P_{\mathrm{g}}(t)$ obtained from Cases 1-3 are applied to the grid dynamic model \eqref{eq:aggregate_grid} to evaluate the grid frequency responses, and the results are illustrated in Figure~\ref{fig:frecom}.

As shown in Figure~\ref{fig:powercom}, 
the grid-side load $P_{\mathrm{g}}(t)$ in Case 2 is smoother than the raw data center demand $P_{\mathrm{dc}}(t)$ but still exhibits relatively large power variations, particularly during the low-load period from approximately 190s to 220s. In contrast, Case 3 yields smaller overall power fluctuations, with $P_{\mathrm{g}}(t)$ maintained mainly within the power envelope of $[120,130]$~MW, similar to Case 1 shown in Figure \ref{fig:power}. 
That is because Case 2 focuses solely on suppressing the power component associated with the targeted grid oscillation mode, while Case 3 focuses on satisfying the power-envelope and ramp-rate requirements.

Nevertheless, Figure \ref{fig:frecom} shows that Case 3 results in larger frequency oscillations than Cases 1 and 2, although all three cases substantially reduce the frequency deviations compared to the baseline case with the raw data center demand $P_{\mathrm{dc}}(t)$. This  indicates that the grid frequency oscillations are driven primarily by the load component near the natural frequency of 0.5~Hz rather than by the overall magnitude of the power variations. Although Case 3 reduces the overall power fluctuations, it does not explicitly suppress the 0.5 Hz oscillatory component. By contrast, Case 2 effectively attenuates this vulnerable frequency component, leading to frequency oscillations comparable to those in Case 1 and substantially smaller than those in Case 3.

\begin{figure}
    \centering
    \includegraphics[width=1\linewidth]{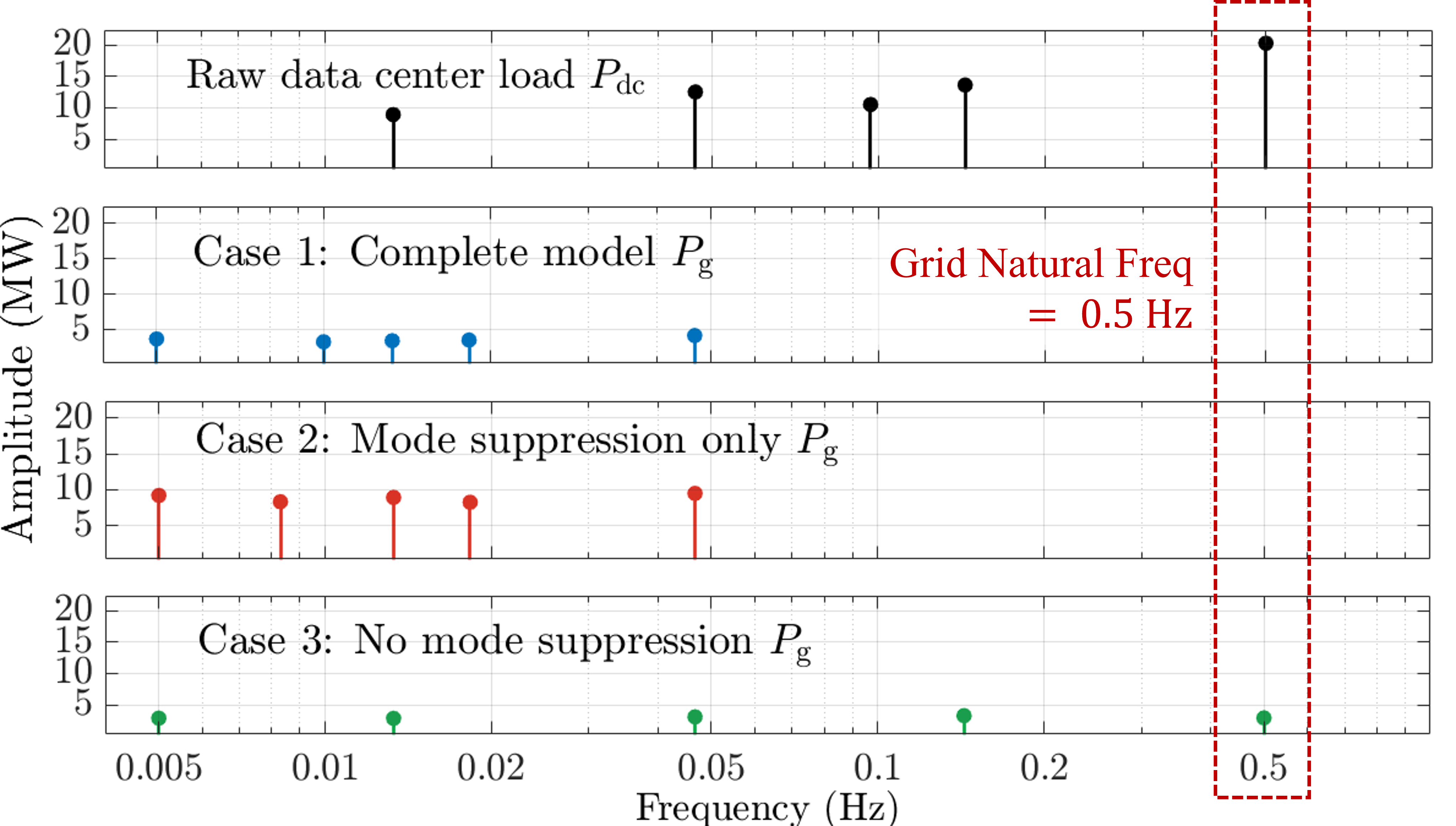}
    \caption{Five dominant frequency components of the raw data center demand $P_{\mathrm{dc}}(t)$ and the grid-side loads $P_{\mathrm{g}}(t)$ in Cases 1-3, obtained via the fast Fourier transform (FFT) analysis.}
    \label{fig:fft}
\end{figure}

To verify this observation, a fast Fourier transform (FFT) analysis is performed on the raw data center demand $P_{\mathrm{dc}}(t)$ and the grid-side loads $P_{\mathrm{g}}(t)$ obtained in Cases 1-3 to identify their top five dominant frequency components. The results are shown  in Figure~\ref{fig:fft}. The raw demand $P_{\mathrm{dc}}(t)$ contains a pronounced component at the grid natural frequency of 0.5~Hz with an amplitude of approximately 20~MW. 
Cases 1 and 2 effectively eliminate this component through the proposed G-MPC algorithm with targeted mode suppression. Although Case 2 retains relatively large low-frequency components, these components have a limited effect on the oscillatory grid response as they are far from the grid natural frequency. In contrast, Case 3 substantially reduces the overall power variations and attenuates the low-frequency components but retains a noticeable component at 0.5~Hz because mode suppression is not considered. Consequently, Case 3 produces larger grid frequency oscillations than Case 2 despite yielding a smoother grid-side load with smaller overall power variations.

This result indicates that suppressing vulnerable oscillatory components in AI data center loads can be more critical for mitigating grid frequency oscillations than merely reducing overall power fluctuations. In addition, the total HESS operational costs over the 300s simulation period are \$99.48, \$72.39, and \$74.17 for Cases 1-3, respectively. Thus, Case 2 incurs the lowest HESS operational cost while achieving grid frequency oscillation mitigation comparable to that of Case 1 and superior to that of Case 3.

\subsection{Terminal Energy Conditions for BESS and SC}
\label{subsec:SOCenforce}

As shown in Figure~\ref{fig:energy}, the stored energy of the BESS gradually increases, indicating net energy accumulation, while the SC also exhibits a deviation from its initial energy level. This behavior occurs because the current G-MPC optimization models do not explicitly regulate terminal energy or SOC deviations. Over extended operation, such energy drift may cause the energy level to approach its upper or lower bound,  reducing the available energy margin for accommodating future load increases or decreases. This issue can be mitigated by introducing a terminal penalty to the optimization models, such as $\rho_E(E_{\mathrm{bat},N}-E_{\mathrm{bat},0})^2$, to discourage deviations from the initial energy level. Alternatively, the hard terminal energy constraints \eqref{eq:E:ini} can be imposed on the BESS and SC to 
restore their terminal energy levels to the initial values, thus ensuring energy neutrality over the optimization horizon.
\begin{align}
     E_{\tb,N} =  E_{\tb,0},\ \  E_{\ts,N} =  E_{\ts,0}.\label{eq:E:ini}
\end{align}

\begin{figure}
    \centering
    \includegraphics[width=1\linewidth]{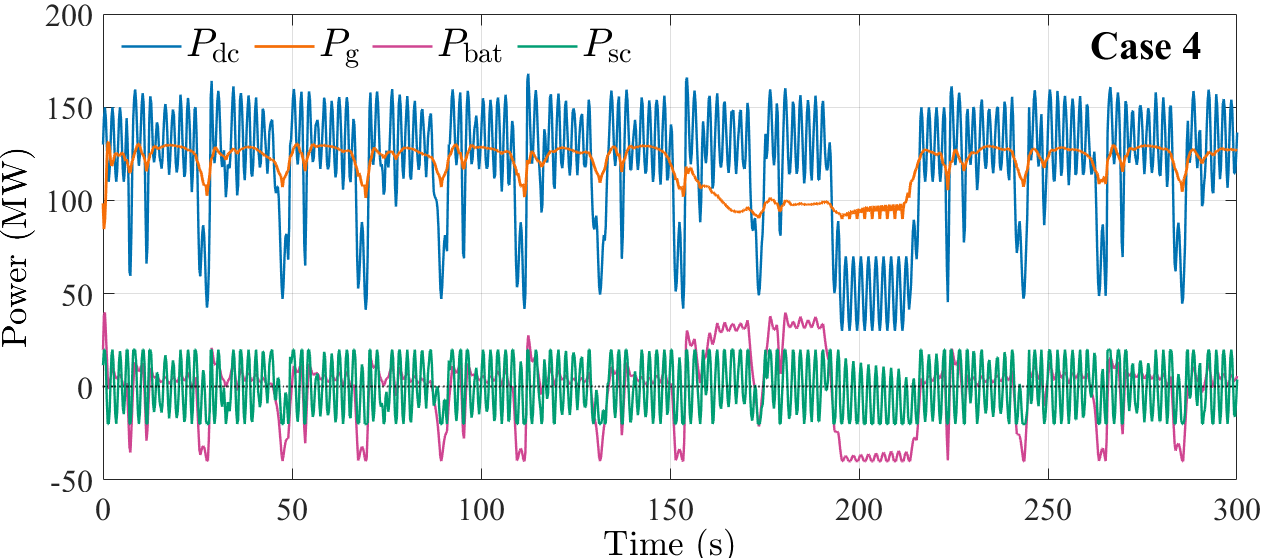}
    \caption{AI data center power demand $P_{\mathrm{dc}}(t)$, grid-side load $P_{\mathrm{g}}(t)$, BESS and SC power outputs $P_\tb(t),P_\ts(t)$ under the proposed G-MPC method in Case 4.}
    \label{fig:case4power}
\end{figure}

\begin{figure}
    \centering
    \includegraphics[width=1\linewidth]{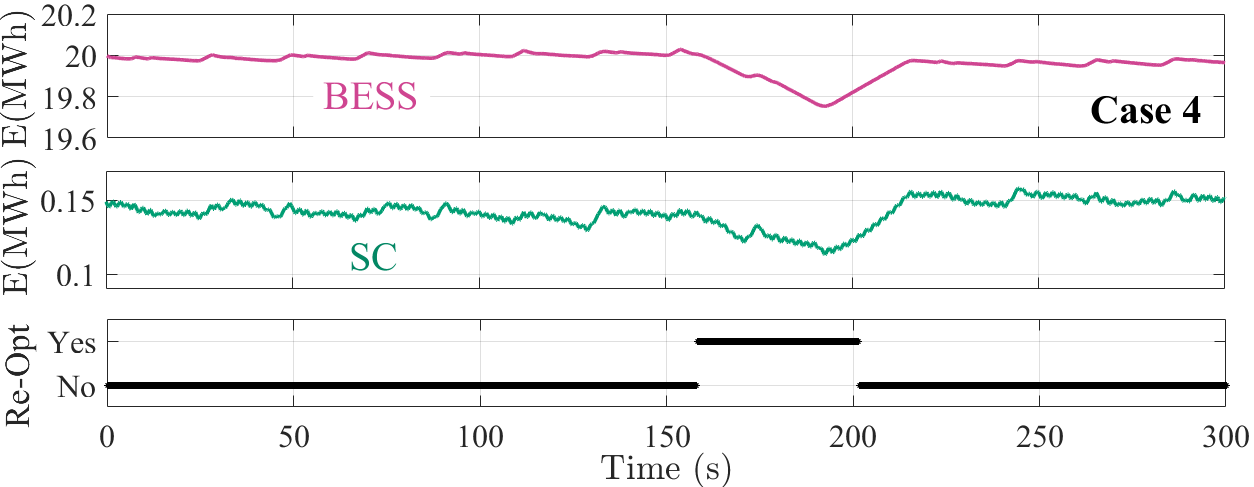}
    \caption{Stored energy $ E_{\tb}(t), E_{\ts}(t)$ of BESS and SC, and re-optimization status under the proposed G-MPC method in Case 4.}
    \label{fig:case4soc}
\end{figure}

Hence, we test the following Case 4 in this subsection:
\begin{itemize}
    \item \emph{Case 4 (Complete Model with Terminal Energy Constraints):} The targeted mode-suppression objective and the grid-side power envelope and ramp-rate requirements are considered in the optimization models, as well as the terminal energy constraints \eqref{eq:E:ini} for the BESS and SC. 
\end{itemize}
Figure~\ref{fig:case4power} presents the resulting grid-side load and the power outputs of the BESS and SC. The corresponding stored-energy trajectories of the BESS and SC, together with the re-optimization status at each MPC step, are shown in Figure~\ref{fig:case4soc}.

Comparing Figure~\ref{fig:case4power} with Figure~\ref{fig:power}, the grid-side load $P_{\mathrm{g}}(t)$ in Case 4 is similar to that in Case 1, except during the interval from approximately 150s to 200s. This difference is primarily caused by the substantial reduction in $P_{\mathrm{dc}}(t)$ from about 190s to 220s. To smooth this load reduction while satisfying the terminal energy constraints \eqref{eq:E:ini}, the BESS and SC discharge in advance and subsequently charge during the low-load period. This behavior demonstrates the anticipatory capability of G-MPC, which adjusts the storage operation in preparation for predicted future load variations. Consequently, as shown in Figure~\ref{fig:case4soc}, the stored-energy levels of both the BESS and SC remain close to their initial values.

\begin{figure}
    \centering
    \includegraphics[width=1\linewidth]{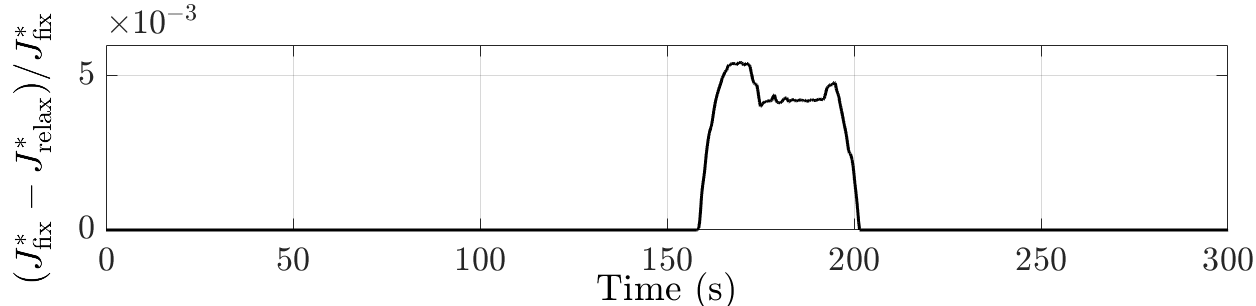}
    \caption{The relative objective value gap $(J^{\star}_{\mathrm{fix}}-J^{\star}_{\mathrm{relax}})/J^{\star}_{\mathrm{fix}}$ in Case 4.}
    \label{fig:gap}
\end{figure}

In addition, Figure~\ref{fig:case4soc} shows that the fix-and-re-optimize step in Algorithm \ref{alg:reopt} is activated from approximately 158s to 201s in the simulation, during which the relaxed model \eqref{eq:relax} and the fixed-status model \eqref{eq:fix} are solved sequentially to obtain the final HESS control decisions. This activation is attributed to the HESS response to the substantial reduction in $P_{\mathrm{dc}}(t)$ while satisfying the terminal energy constraints \eqref{eq:E:ini}. Nevertheless, as shown in Figure~\ref{fig:gap}, the relative objective value gap, defined as $(J^{\star}_{\mathrm{fix}}-J^{\star}_{\mathrm{relax}})/J^{\star}_{\mathrm{fix}}$, remains very small with a maximum value of $0.55\%$ and an average value of $0.4\%$ across the re-optimized instances. This small gap indicates the (near-)optimality of the final control decisions generated by the proposed Algorithm~\ref{alg:reopt}.

\subsection{Computational Efficiency}\label{subsec:efficiency}

\begin{figure}[H]
    \centering
    \includegraphics[width=1\linewidth]{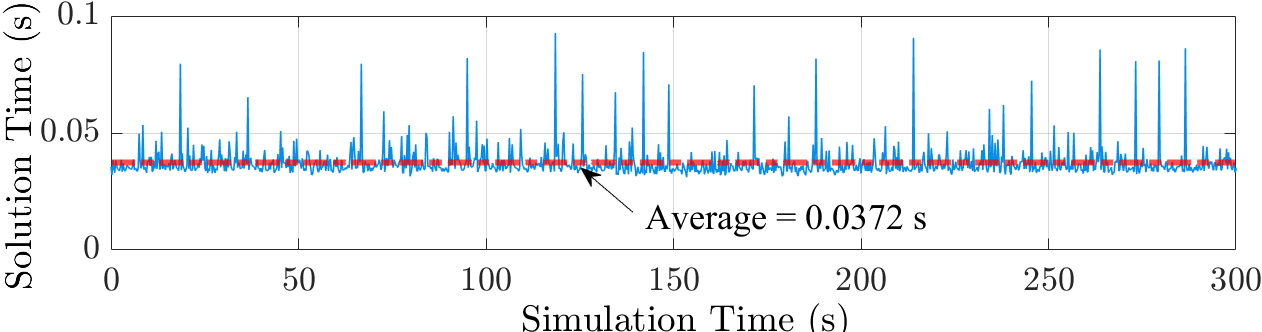}
    \caption{Solution time of the complete G-MPC optimization model at each time step in Case 1.}
    \label{fig:solvetime}
\end{figure}

The numerical experiments are conducted on a computer with an Intel(R) Core(TM) i7-1185G7 CPU operating at 3.00~GHz and 16~GB of RAM. The G-MPC optimization models are formulated in Julia using the JuMP package \cite{Lubin2023} and solved with Gurobi Optimizer 10.0.1 \cite{gurobi}. For the complete model in Case 1, the solution time at each step is shown in Figure~\ref{fig:solvetime}. The average solution time is 0.0372~s, and all G-MPC models are solved within 0.1~s. These results demonstrate the computational efficiency of the proposed method and its suitability for real-time control implementation.

\section{Conclusion}
\label{sec:conclusion}

In this paper, we develop a grid-mode-aware model predictive control (G-MPC) framework for coordinating a BESS and an SC within an HESS to smooth the grid-side power demand of AI data centers. By directly embedding band-pass filter dynamics into the optimization model, the proposed framework explicitly extracts and suppresses power components associated with vulnerable grid oscillatory modes. The G-MPC formulation jointly considers grid-side power smoothing requirements and HESS operational costs, while satisfying physical operating constraints. A computationally efficient fix-and-re-optimize algorithm is developed to prevent simultaneous charging and discharging while requiring at most two convex QP solutions at each control step.
Simulation results demonstrate that the proposed framework effectively smooths highly variable AI data center demand, suppresses vulnerable power components, and coordinates the complementary power and energy capabilities of the BESS and SC. The ablation studies further show that merely reducing overall power variations does not necessarily mitigate grid oscillations, underscoring the importance of explicitly suppressing power components associated with vulnerable grid modes. Future work will incorporate advanced AI data center load prediction techniques and account for prediction uncertainty through robust or stochastic MPC formulations.

\bibliographystyle{IEEEtran}
\bibliography{IEEEabrv,mybibfile}

\end{document}